\documentclass[twocolumn,pra,amsmath,amssymb,superscriptaddress, longbibliography]{revtex4-1}
\usepackage{graphicx,bm,dsfont,amsmath,amssymb}
\usepackage{soul}
\usepackage{color}
\usepackage{amsthm}
\usepackage{mathrsfs}
\usepackage{subfigure}
\usepackage{lipsum}
\usepackage{float}
\usepackage[colorlinks=true,citecolor=blue,urlcolor=blue]{hyperref}
\normalsize
\usepackage[percent]{overpic}
\usepackage[ruled,vlined, linesnumbered]{algorithm2e}

\begin{document}

\title{Speedup Chip Yield Analysis by Improved Quantum Bayesian Inference}
\author{Zi-Ming Li}
\affiliation{School of Integrated Circuits, Tsinghua University, Beijing 100084, China}
\author{Zeji Li}
\affiliation{School of Integrated Circuits, Tsinghua University, Beijing 100084, China}
\author{Tie-Fu Li}
\affiliation{School of Integrated Circuits, Tsinghua University, Beijing 100084, China}
\affiliation{Frontier Science Center for Quantum Information, Beijing, China}
\author{Yu-xi Liu}
\email{yuxiliu@mail.tsinghua.edu.cn}
\affiliation{School of Integrated Circuits, Tsinghua University, Beijing 100084, China}
\affiliation{Frontier Science Center for Quantum Information, Beijing, China}

\date{\today}

\date{\today}

\begin{abstract}
The semiconductor chip manufacturing process is complex and lengthy, and potential errors arise at every stage. Each wafer contains numerous chips, and wafer bin maps can be generated after chip testing. By analyzing the defect patterns on these wafer bin maps, the steps in the manufacturing process where errors occurred can be inferred. In this letter, we propose an improved quantum Bayesian inference to accelerate the identification of error patterns on wafer bin maps, thereby assisting in chip yield analysis. We outline the algorithm for error identification and detail the implementation of improved quantum Bayesian inference. Our results demonstrate the speed advantage of quantum computation over classical algorithms with a real-world problem, highlighting the practical significance of quantum computation.
\end{abstract}

\maketitle
\textit{Introduction.---}Semiconductor manufacturing processes are lengthy and intricate, often comprising hundreds of steps. Each of these steps has the potential for errors that can eventually result in defects in the chips produced on the wafer~\cite{plummer2009silicon, may2006fundamentals, quirk2001semiconductor,franssila2010introduction}. The yield of chip production refers to the proportion of total chips that meet quality standards. Chip yield is a critical metric in the semiconductor industry, its improvement can lead to advanced production efficiency and reduced costs~\cite{stapper1995integrated, kumar2006review, cunningham1995semiconductor}.

To improve chip yield, it is essential to identify the steps in the manufacturing process that can contribute to chip defects and optimize those specific steps. Each wafer contains a substantial number of chips, and once the manufacturing process is complete, each chip undergoes testing. Qualified chips are recorded as zero, while defective chips are marked as one, resulting in a wafer bin map (WBM)~\cite{plummer2009silicon, may2006fundamentals,quirk2001semiconductor,franssila2010introduction}. Errant chips on the wafer often do not occur independently; instead, they exhibit correlations with errors on other chips. These errors form specific collective error patterns on the wafer bin maps, which contain information about the steps in the manufacturing process where errors occurred. By analyzing the error patterns on the WBM, it is possible to infer the problematic steps in the process, helping engineers to refine the chip manufacturing process, ultimately increasing the chip production yield~\cite{chien2007data, hsu2007hybrid, chao2010constructing, liao2013similarity}

In chips mass production, the manpower and costs associated with manually identifying error patterns on wafers are increasing, and the accuracy of such methods cannot be guaranteed. Therefore, there is a need for automatic recognition algorithms. Numerous algorithms based on machine learning and deep learning have been proposed to address this challenge, achieving automatic recognition of the pattern of errors~\cite{wang2020deformable, wang2020defect, kim2023advances, ishida2019deep, chien2007data}. However, the speed or the accuracy of these methods is limited, and they struggle to handle cases with partially missing data effectively, which frequently occurs in real industrial situations. There is an urgent need for a classification method that combines high speed and accuracy while effectively handling the partially missing data cases.

With the introduction of the concept of quantum computation, many quantum algorithms have been proposed to achieve speedups over classical algorithms\cite{shor1994algorithms, shor1999polynomial, grover1996fast, brassard2002quantum,harrow2009quantum,childs2017quantum,rebentrost2014quantum,lloyd2014quantum,amin2018quantum,prakash2014quantum}. Quantum Bayesian inference (QBI) algorithm on the quantum Bayesian network is one of them. The concept of a quantum Bayesian network is first proposed as a quantum analogue of the classical Bayesian network~\cite{tucci1997quantum}. Subsequent work has proposed quantum circuit representations of classical Bayesian networks and quantum Bayesian inference~\cite{low2014quantum, borujeni2021quantum, fathallah2023optimized}. The previous quantum Bayesian inference algorithm provides a quadratic speedup over its classical counterparts~\cite{low2014quantum}. But its speedup is limited and an end-to-end procedure has not been explicitly proposed for performing quantum Bayesian inference on general Bayesian networks.

In this letter, we present an improved quantum Bayesian inference algorithm that achieves a quadratic speedup over both previous quantum and classical Bayesian inference algorithms. An end-to-end procedure is proposed for performing quantum Bayesian inference. We employ the proposed method to identify and classify specific error patterns on wafer bin maps. We demonstrate that the proposed algorithm effectively identifies wafer error patterns with nearly optimal accuracy, and is able to handle the partially missing data cases.

\begin{figure*}[t]
	\centering
	\begin{overpic}[scale=0.45]{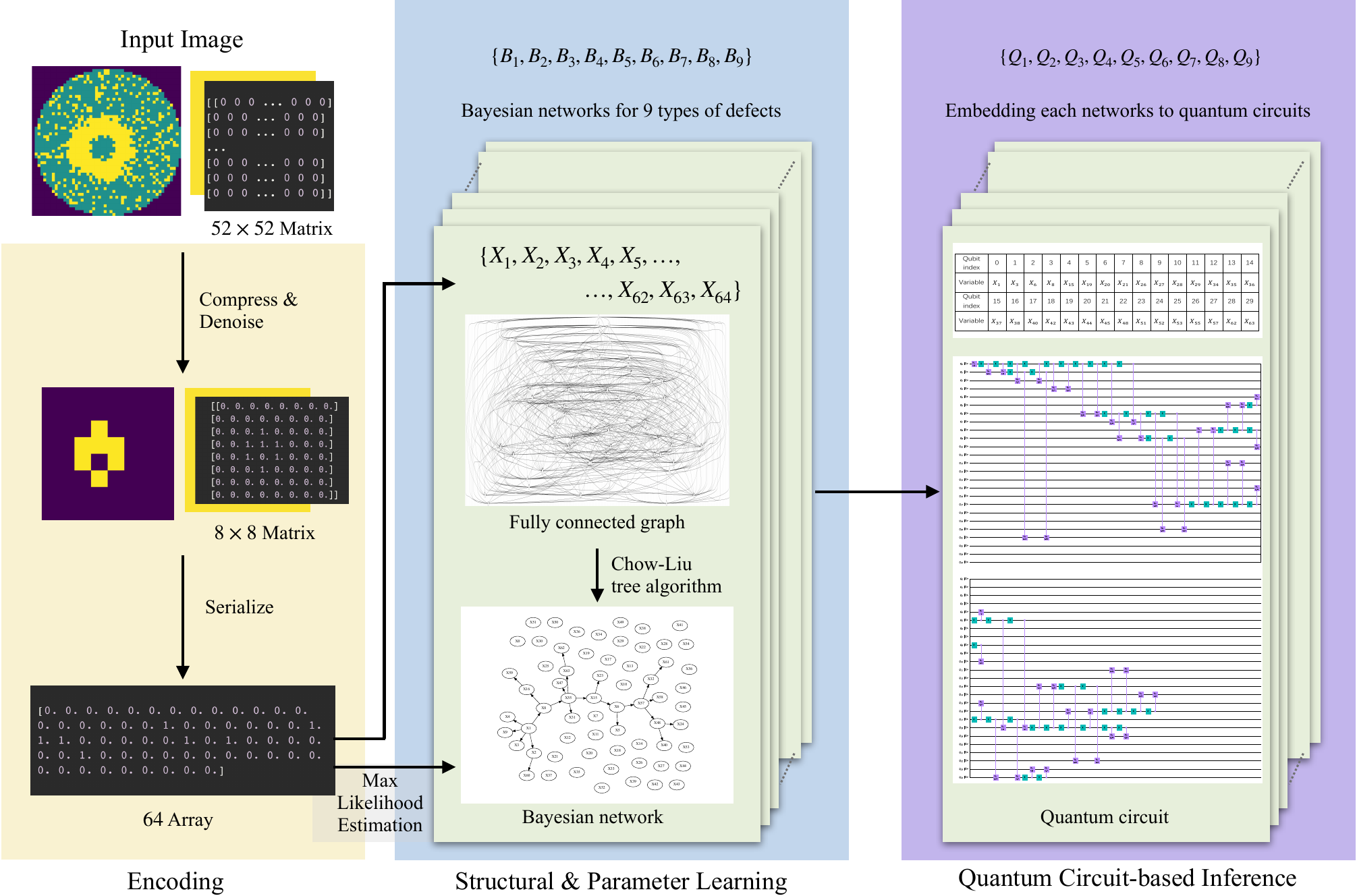}
		\put(0,63.5){\small\textbf{(a)}}
		\put(0,39){\small\textbf{(b)}}
		\put(0,16.5){\small\textbf{(c)}}
		\put(51,40.8){\small\textbf{(d)}}
		\put(51,19.5){\small\textbf{(e)}}
		\put(69.5,48){\small\textbf{(f)}}
	\end{overpic}
	\caption{(a)-(e) Constructing nine Bayesian networks $\{\mathcal{B}_1,\cdots,\mathcal{B}_9\}$ according to the data of nine types of defect. (f) Constructing nine quantum circuits $\{\mathcal{Q}_1,\cdots,\mathcal{Q}_9\}$ encoding the nine Bayesian networks $\{\mathcal{B}_1,\cdots,\mathcal{B}_9\}$.}
	\label{fig1}
\end{figure*}
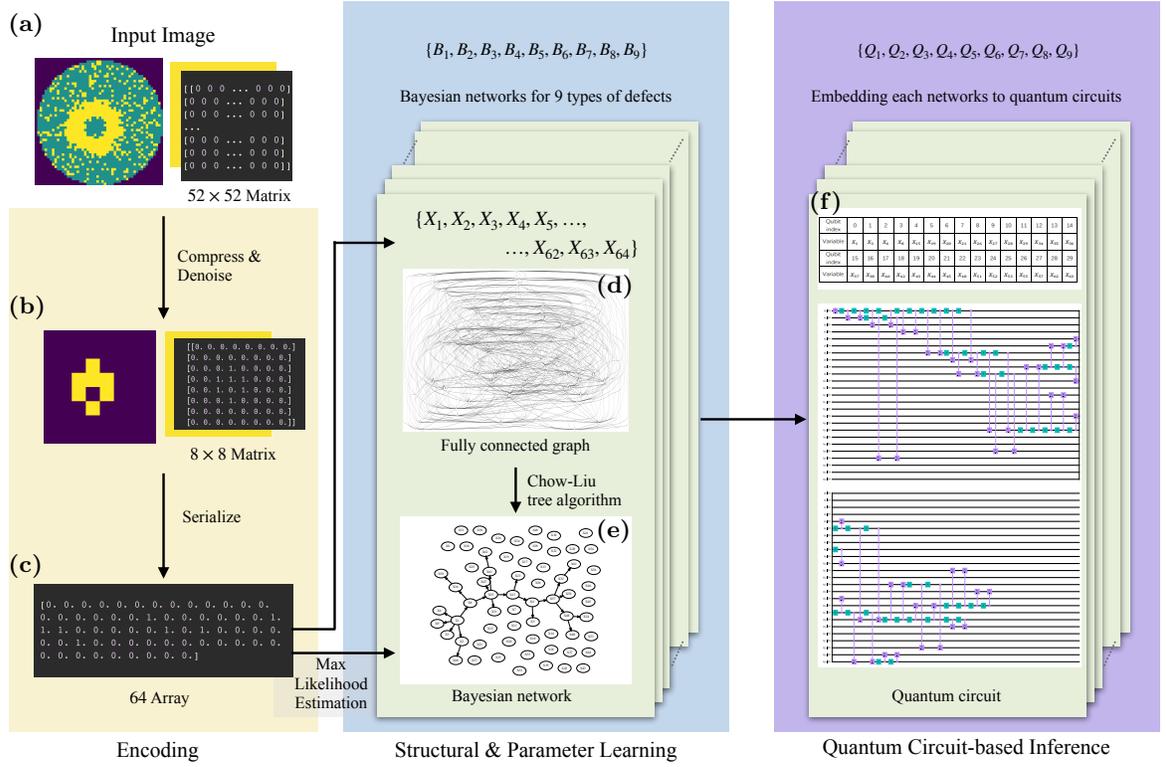

\textit{Bayesian networks based on chip yield analysis.---} There are many problems that can be dealt by Bayesian network theory. For concreteness of discussions and as an example, we here show how to construct Bayesian networks based on an open defect data set in wafer bin maps~\cite{wang2020deformable}. There are nine single-type defects in this open dataset: Normal, Center, Doughnut, Edge-Loc, Edge-Ring, Loc, Near-Full, Scratch and Random. For each type of defect, the dataset includes 1000 wafer bin maps where the data is acquired from a wafer manufacturing facility. Some of the dataset examples are shown in Appendix.~\ref{ap9}.

As shown in Fig.~\ref{fig1}(a), we represent each wafer bin map with one type of defect as a $52$-by-$52$ two-dimensional array consisting only zero and one, where zero and one indicate a chip without and with defects, respectively. There are also wafer bin maps containing multi-type defects in this dataset, but in this study we only identify and classify the wafer bin maps with single-type defects. We compress each of the 52-by-52 wafer bin maps to an $8$-by-$8$ two-dimensional array as shown in Fig.~\ref{fig1}(b), then flatten each of the $8$-by-$8$ two-dimensional array to a one-dimensional array with length 64 as shown in Fig.~\ref{fig1}(c). This process filters out random noise, reduces the size of the input data, and emphasizes the key features of the error patterns. We find that the minimum size of the compressed data as in Fig.~\ref{fig1}(b) is \(8 \times 8\). If the data is compressed to a smaller size, the primary features of the error pattern will be lost, leading to a significant decline in classification accuracy. We detail the compression procedure and give some examples in Appendix.~\ref{ap9}.

We denote the 64 elements of each one-dimensional array as 64 random variables $X\equiv\{X_1,\cdots,X_{64}\}$ as shown in Fig.~\ref{fig1}(d) with $X_k=0$ or $1$, $\forall \,\, k=1,\cdots,64$. For $i$th type of defect with $i=1,\cdots,9$, we find a Bayesian network $\mathcal{B}_i$ with 64 nodes  $\mathcal{N}\equiv\{\mathcal{N}_1,\cdots,\mathcal{N}_{64}\}$ using the Chow-Liu tree structure learning algorithm~\cite{chow1968approximating} to express the correlations between the 64 random variables.  The $k$th node is assigned to correspond  random variable $X_k$.  Thus, nine types of Bayesian networks $\mathcal{B}\equiv \{\mathcal{B}_1,\cdots,\mathcal{B}_9\}$ are constructed, and the structure of each Bayesian network is schematically shown in Fig~\ref{fig1}(e). Each $\mathcal{B}_i$ is constructed with maximum indegree $1$ to facilitate the subsequent construction of quantum circuits. The whole Chow-Liu procedure is explained in Appendix.~\ref{ap2}. We finally determine the conditional probability table for nodes of each Bayesian network based on the data in arrays as shown in Fig.~\ref{fig1}(c) with the maximum likelihood estimation. Thus, the joint probability distributions $P(X_1,\cdots, X_{64})$ of $64$ random variables and structures of Bayesian networks corresponding to nine types of defects are given.

\textit{Quantum circuit representation of classical Bayesian networks.} To implement quantum computation of classical Bayesian inference, we use 64 qubits to represent 64 nodes of each constructed Bayesian network $\mathcal{B}_i$, and two computational basis states $|q_{k}=0\rangle$ and  $|q_{k}=1\rangle$ of the $k$th qubit  denote two values of random variable $X_k$ at the $k$th node. All of node probabilities $P(X_1,\cdots, X_{64})$  in the conditional probability table of a Bayesian network $\mathcal{B}_i$  are encoded as amplitudes of the quantum state
\begin{equation}\label{eq2}
	|\psi_i\rangle=\sum_{q_1,\dots,q_{64}}\sqrt{P(X_1=q_1,\cdots, X_{64}=q_{64})}|q_1\cdots q_{64}\rangle.
\end{equation}
in qubit computational basis states. The encoded quantum state $|\psi_i\rangle$  is realized via a quantum circuit $\mathcal{Q}_i $ acting on the initial state $|00\cdots 0\rangle$ of the 64 qubits as schematically shown in Fig.~\ref{fig1}(f).  Thus,  nine Bayesian networks $\{\mathcal{B}_1,\cdots,\mathcal{B}_9\}$ are represented by nine quantum states $\{|\psi_1\rangle,\cdots,|\psi_9\rangle\}$.

Encoding the probabilities $P(X_1,\cdots, X_{64})$ to the quantum state $|\psi_i\rangle$ via quantum circuit $\mathcal{Q}_i $ requires several steps. We need first extract a directed acyclic graph $\mathcal{G}_i=\langle \mathcal{N},\mathcal{E}_i\rangle$ from $\mathcal{B}_i$ and perform the topological sorting algorithm~\cite{kahn1962topological}, with the sets $ \mathcal{N}$ and  $\mathcal{E}_i$ of nodes  and edges.  That is, all nodes in the node set $ \mathcal{N}$ are rearranged into a new node set $T$ so that the nodes $T_1,\cdots, T_{64}$  in $T$ are sorted topologically from upstream to downstream.  Each node $T_k$ corresponds to a random variable $Z_k$ satisfying $Z_k\in X$ and the indices of the parents of $T_k$ is smaller than $k$, $\forall\,\, k=1,2,\cdots,64$. Thus, each node or edge is visited once in the topological sorting, so the complexity is $O(|\mathcal{N}|,|\mathcal{E}_i|)$ with the numbers $|\mathcal{N}|$ and $|\mathcal{E}_i|$ of the nodes and edges. The topological sorting algorithm is detailed in Appendix.~\ref{ap3}.

Following the topological sorting, the $64$-qubit quantum state $|\psi_i\rangle$ can be generated recursively. That is, we first generate a single-qubit state $|\psi_{i,1}\rangle$, which encodes the variable $Z_1$  at the node $T_1$ for two values $Z_1=0$ and $Z_1=1$  with their probabilities. This can be done by initializing a qubit representing $Z_1$ in the state $|0\rangle$ and following up  a rotation $\rm{R_Y}(\theta)=\exp(-i\theta\sigma_y/2)$ around the $y$-axix with $\theta=2\arccos\sqrt{P(Z_1=0)}$, such that
\begin{equation}
	|\psi_{i,1}\rangle=\sqrt{P(Z_1=0)}|0\rangle+\sqrt{P(Z_1=1)}|1\rangle.
\end{equation}
Suppose a quantum state $|\psi_{i,k}\rangle$ of $k$ qubits has been generated from the initial state $|0\cdots 0\rangle$. That is, we use probability amplitudes of the state
\begin{equation}
	|\psi_{i,k}\rangle=\sum_{q_1,\dots,q_{k}}\sqrt{P(Z_1=q_1,\cdots,Z_k=q_k)}|q_1q_2\cdots q_{k}\rangle
\end{equation}
to represent distribution of $2^k$ joint probabilities for $k$ nodes characterized by $k$ random variables $\{Z_1,\dots,Z_k\}$, where the states of $k$th qubit represent the values of variable $Z_k$. We aim to construct state $|\psi_{i,k+1}\rangle$, which represents the joint  probability distribution of $k+1$ nodes with variables $\{Z_1,\dots,Z_{k+1}\}$. Based on the topological sorting property,  $T_{k+1}$ is a downstream node for $\{T_1,T_2,\dots,T_k\}$, thus the set $pa(T_{k+1})$ of parent nods for $T_{k+1}$ satisfies $pa(T_{k+1})\subset \{T_1,T_2,\dots,T_k\}$. The probabilities $P(Z_1,Z_2,\cdots, Z_{k+1})$ for $k+1$ variables are
\begin{equation}
	\label{eq3}
	P(Z_1,Z_2,\cdots, Z_{k+1})=P(Z_{k+1}|pa(Z_{k+1}))P(Z_1,\cdots, Z_k).
\end{equation}
The $(k+1)$th qubit is initialized in state $|0\rangle$, then controlled-Y operation $\rm{CR_Y}(\theta)$ are applied to the $(k+1)$th qubit such that the state $|q_1q_2\cdots q_k0\rangle$ is turned into
\begin{equation}
	\label{eq4}
	\mathrm{CR_Y}
	(\theta)|q_1q_2\cdots q_k0\rangle=|q_1q_2\cdots q_k\rangle\left(\cos\frac{\theta}{2}|0\rangle+\sin\frac{\theta}{2}|1\rangle\right),
\end{equation}
$\forall q_1,q_2,\cdots q_k=0,1$, with the parameter $\theta$ determined by $P(Z_{k+1}|pa(Z_{k+1}))$ in Eq.~(\ref{eq3}). The control qubits are the qubits which represent parent variables $pa(Z_{k+1})$ and the controlled qubit is the $(k+1)$th qubit. Suppose the number of parent nodes of $T_{k+1}$ is $m_{k+1}$, a total number of $2^{m_{k+1}}$ controlled-Y operations are needed to construct $|\psi_{i,k+1}\rangle$. These controlled operations can be realized with the methods in Refs.~\cite{bergholm2005quantum, allcock2023constant}.  Thus $O(2^{m_{k+1}})$ quantum operations are used to construct $|\psi_{i,k+1}\rangle$ from $|\psi_{i,k}\rangle$. Following this procedure, the target state $|\psi_i\rangle$ can be generated. Thus, the quantum circuit $\mathcal{Q}_i$ in Fig.~\ref{fig1}(f) corresponding to $|\psi_i\rangle$ is constructed by $\rm{R_{Y}}(\theta)$ operation and controlled $\rm{R_{Y}}(\theta)$ operations. Generally, if we denote $N$ the variable number and $m$ the maximum indegree of a Bayesian network, a total number of $O(N2^m)$ quantum operations are needed to construct a state encoding the Bayesian network from the initial state $|00\dots0\rangle$. In our case, we have $N=64$ and $m=1$, thus only a constant number of quantum operations are needed to construct each $|\psi_i\rangle$. With the above procedure,  the main feature of each type of defects is encoded in the corresponding quantum state. We summarize the procedure in Appendix.~\ref{ap3} and prove it rigorously in Appendix.~\ref{proof2}. 

It is noted that the complexity of $O(N2^m) $is exponential in parameter $m$, thus we provide an optimization method for reducing the maximum indegree in Appendix.~\ref{ap4}. Given a Bayesian network, we can perform the pre-process algorithm as in Appendix.~\ref{ap4}, thus maximizing the advantages of quantum algorithms and avoiding the high complexity the quantum state construction procedure.

Below, we show how to classify wafer bin maps based on the constructed quantum state by performing quantum Bayesian inference.

\textit{Improved quantum Bayesian inference.---} Let us now propose an improved quantum Bayesian inference (QBI) algorithm for chip yield analysis. In contrast to perform quantum amplitude amplification algorithm and direct sampling as in~\cite{low2014quantum}, we employ the  quantum amplitude estimation algorithm~\cite{brassard2002quantum,montanaro2015quantum,giurgica2022low,grinko2021iterative} and improve it as follows.

The quantum amplitude estimation algorithm with arbitrary success probability takes as input one copy of a quantum state $|\psi\rangle$, a unitary transformation $U=2|\psi\rangle\langle \psi|-I$, another unitary transformation $V=I-2P$ for some projector $P$, an error rate $\epsilon$, and probability of failure $\delta$. The the algorithm output is $\tilde{a}$, which  is an estimate of $a = \langle\psi|P|\psi\rangle$. If $ |\tilde{a}-a|/|a|\leq \epsilon$
with success probability at least $1-\delta$, then  $O\left(\ln(2/\delta)/\epsilon\sqrt{a}\right)$ times $UV$ are required.  If $a=0$ or $a=1$ then $\tilde{a}=a$ with certainty. We summarize our conclusion as a theorem in Appendix.~\ref{ap3} and give our proof in Appendix.~\ref{proof1}. The original quantum amplitude estimation algorithm~\cite{brassard2002quantum} is detailed in Appendix.~\ref{ap6}.

We now show how to use improved quantum amplitude estimation on the state $|\psi \rangle$ encoding a Bayesian network $\mathcal{B}$ to perform QBI for  $P(\mathcal{Y}|\mathcal{X}=x)$  when evidence variable set $\mathcal{X}=x$ and target variable set $\mathcal{Y}$ are given.  Taking the projector $P$ as
\begin{equation}
	P=I\otimes |\mathcal{X}=x\rangle \langle \mathcal{X}=x|,
\end{equation}
where the state $|\mathcal{X}=x\rangle$ encodes the  evidence variable $\mathcal{X}$ with $\mathcal{X}=x$. Thus, the value of
$\langle \psi|P|\psi \rangle=P(\mathcal{X}=x)$ can be estimated by quantum amplitude estimation. Similarly, for each possible value $y$ of $\mathcal{Y}$, the value of $P(\mathcal{Y}=y,\mathcal{X}=x)$ with $P^\prime=I\otimes |\mathcal{X}=x,\mathcal{Y}=y\rangle\langle \mathcal{Y}=y, \mathcal{X}=x|$ can be estimated. Thus, following the Bayesian formula
\begin{equation}
	P(\mathcal{Y}=y|\mathcal{X}=x)=\frac{P(\mathcal{Y}=y,\mathcal{X}=x)}{P(\mathcal{X}=x)},
\end{equation}
the conditional probability distribution $P(\mathcal{Y}|\mathcal{X}=x)$ can be estimated with arbitrary precision and probability.

In our algorithm, the input is the state  $|\psi\rangle$, the sets of $\mathcal{X}$ and $\mathcal{Y}$, the value of evidence variables $\mathcal{X}=x$, an error rate $\epsilon$, and probability of failure $\delta$. If the algorithm output $\overline{P}(\mathcal{Y}|\mathcal{X}=x)$, which is an estimate of $P(\mathcal{Y}|\mathcal{X}=x)$,  satisfies
\begin{equation}
	\frac{|\overline{P}(\mathcal{Y}=y|\mathcal{X}=x)-P(\mathcal{Y}=y|\mathcal{X}=x)|}{P(\mathcal{Y}=y|\mathcal{X}=x)}\leq \epsilon
\end{equation}
with probability at least $1-\delta$  for arbitrary possible value $y$ of $\mathcal{Y}$, then the complexity of quantum operations to realized this algorithm is
\begin{equation}
	O\left(N2^mP(\mathcal{X}=x)^{-\frac{1}{2}}\frac{(2\sqrt{2})^{|\mathcal{Y}|}}{\epsilon}\ln\frac{2}{\delta}\right),
\end{equation}
in contrast to complexities of classical Bayesian inference $O\left(Nm \ln(2/\delta) 4^{|\mathcal{Y}|}/(\epsilon^2 P(\mathcal{X}=x))\right)$ and  the previous best quantum algorithm~\cite{low2014quantum} $O\left(N2^m \ln(2/\delta) 4^{|\mathcal{Y}|}/(\epsilon^2 \sqrt{P(\mathcal{X}=x)})\right)$, where $N$ is the variable number and $m$ is the maximum indegree of the Bayesian network.  The result is summarized in Appendix.~\ref{ap3} and proved in Appendix.~\ref{proof1}. The previously best QBI algorithm~\cite{low2014quantum} is introduced in Appendix.~\ref{ap5}. It is clear that our result achieves a polynomial speedup on the parameter $1/\epsilon$ and $|\mathcal{Y}|$.

\textit{Chip yield analysis based on improved QBI algorithm.---} We now apply our algorithm to the error pattern classification for open data~\cite{wang2020deformable} by replacing the state $|\psi\rangle$ in the improved QBI algorithm with the states $|\psi_i\rangle$ given in Eq.~(\ref{eq2}). For a target variables characterized by specific data of variables $X$ of a wafer map, the QBI algorithm is performed on the quantum states $\{|\psi_1\rangle,\cdots,|\psi_9\rangle\}$ to find which type of defect does $X$ belongs to with the maximum probability. For each $X$ with $X\equiv \{X_1=x_1,\cdots,X_{64}=x_{64}\}$, where $x_k=0,1$ with $ k=1,\cdots,64$, we denote the probability that $X$ belongs to defect type $i$ as $P\left(\mathcal{C}_i|X\right)$. Then $P\left(\mathcal{C}_i|X\right)$ is calculated with Bayesian formula,
\begin{equation}
	P\left(\mathcal{C}_i|X\right)=\frac{P\left(X|\mathcal{C}_i\right)P\left(\mathcal{C}_i\right)}{P(X)}, \;i=1,\cdots,9.
\end{equation}
Each of the $P\left(X|\mathcal{C}_i\right)$ is encoded in the amplitude of each $|\psi_i\rangle$ as
\begin{equation}
	P\left(X|\mathcal{C}_i\right)=\;\parallel \langle x_1\cdots x_{64}|\psi_i\rangle \parallel^2,  i=1,\cdots,9.
\end{equation}
Thus, each of the conditional probability $P\left(X|\mathcal{C}_i\right)$ can be calculated via performing the proposed QBI algorithm on each $|\psi_i\rangle$. Therefore, we can classify the wafer map $X$ to defect type $i^{*}$ with
\begin{align}
	i^{*}&=\arg\max_i P\left(\mathcal{C}_i|X\right)\nonumber\\
	&=\arg\max_i P\left(X|\mathcal{C}_i\right)P\left(\mathcal{C}_i\right),\;i=1,\cdots,9.
\end{align}
The prior distribution $P\left(\mathcal{C}_i\right)$ is chosen based on real data in industries. The task of error pattern classification on wafer bin maps is completed.

We further show the result of the wafer error pattern classification for open defect data set~\cite{wang2020deformable} by employing our  QBI algorithm. In our simulation, quantum states $|\psi_i\rangle$ are derived using the training set taken from 80$\%$ of total open data set. Subsequently, the QBI algorithm is executed to classify both the training and test sets with  20$\%$ of total open data set. As Bayesian networks favour unsupervised learning approaches, a validation set is unnecessary. The accuracy of the algorithm is determined by measuring the percentage of labels correctly predicted out of the total open data. We find that our QBI algorithm attains a classification accuracy of 97.0$\%$ and 93.6$\%$ in the training and test sets for open defect data set~\cite{wang2020deformable}, respectively. We mention that the QBI algorithm is implemented and simulated with Qiskit noiseless quantum simulator~\cite{Qiskit2021}.

\begin{figure}[t]
	\includegraphics[width=.9\linewidth]{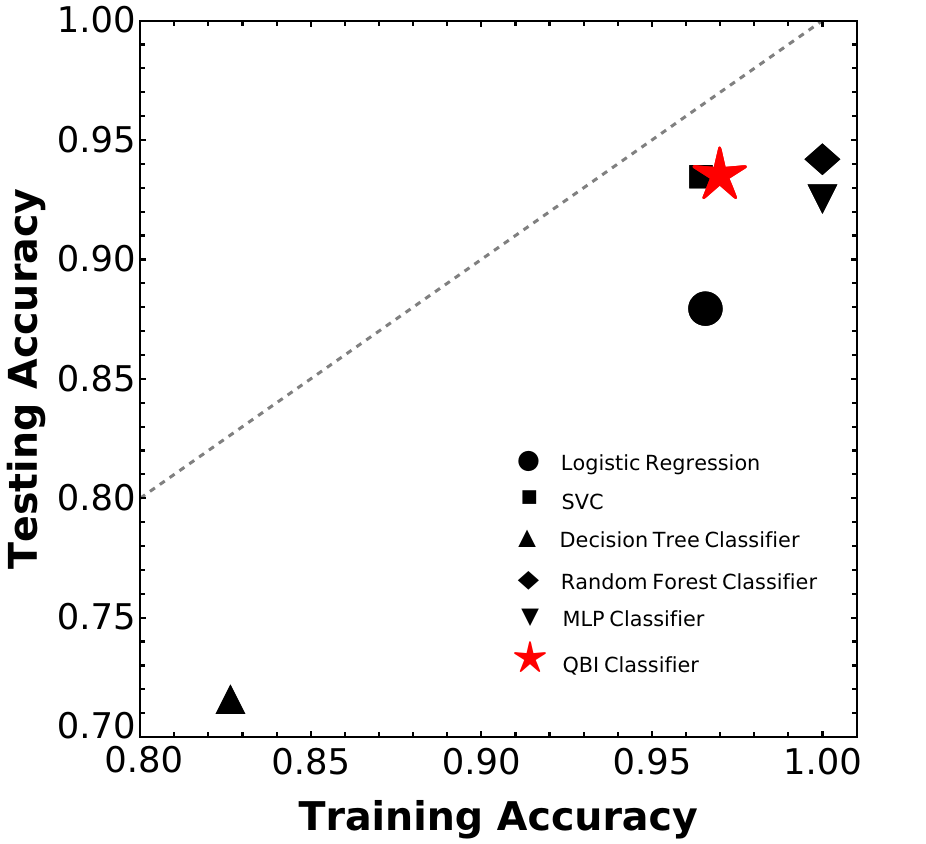}
	\caption{Testing accuracy versus training accuracy for various classification algorithms applied to the dataset. Each point represents a classifier: Logistic Regression (circle), Support Vector Classifier (square), Decision Tree Classifier (upper triangle), Random Forest Classifier (dice), Multi-Layer Perceptron Classifier (lower triangle), and our proposed Quantum Bayesian Inference (QBI) Classifier (star).}
	\label{algo_comparison}
\end{figure}

We compare our results with various classical machine learning methods on this problem, and summarize the results in Fig.~\ref{algo_comparison}. The QBI classifier demonstrates high accuracy comparable to traditional classifiers, indicating its potential effectiveness. Also, the dashed line shown in Fig.~\ref{algo_comparison} corresponds to equal training and testing accuracy, which highlights the generalization performance of each classifier, with points closer to the dashed line suggesting minimal over-fitting. Thus it is shown that the QBI classifier exhibits minimal over-fitting, resulting in strong generalization performance.

\textit{Discussions and Conclusion.---} We propose an improved QBI algorithm based on the amplitude estimation. Compared to the previous best QBI algorithm~\cite{low2014quantum} which has a quadratic speedup over classical one, our algorithm achieves a polynomial speedup.  We use chip yield analysis in semiconductor manufacturing as a concrete example to validate our algorithm.  That is, we convert  wafer error pattern classification into the problem of QBI. Our algorithm achieves a classification accuracy of over 93$\%$ on an open dataset. The classification accuracy based on our QBI algorithm is nearly optimal for this dataset, comparable to the best classical algorithms, while the computational time required by the quantum algorithm is significantly lower than that of classical Bayesian inference running on classical computers. We also note that our discussions are limited to the situation that the variable at each node only takes two values for yield analysis. For the case that a variable has $j$ values with $j>2$, we just use $\lceil \log_2 j\rceil$ qubits to represent one variable, while the state construction and the QBI procedure are the same as the case that each variable has two values.

We emphasize that our algorithm exhibits the capability of handling partially missing data, which is a significant challenge for classical algorithms. In practical chip manufacturing processes, for instance, some values in variables  as previously defined may be missed. When the value for a variable  is missed, we only need to trace out the qubit representing variable and perform the QBI algorithm on the remaining qubits. Thus, our QBI algorithm can handle the cases of partially missing data. Finally, we point out that our study can be applied to not only the chip yield analysis but also the other problems related to classical Bayesian networks, e.g., computational biology and bioinformatics~\cite{needham2007primer}, gene regulatory networks~\cite{zou2005new}, protein structure~\cite{bradford2006insights}, gene expression analysis~\cite{murphy1999modelling}, medical diagnosis~\cite{nikovski2000constructing}, document classification~\cite{denoyer2004bayesian}, information retrieval~\cite{fung1995applying}, decision support systems~\cite{stassopoulou1998application}, image processing~\cite{ho2012wavelet} and artificial intelligence~\cite{korb2010bayesian, niedermayer2008introduction, friedman1996building}.

\begin{acknowledgments}
This work was supported by Innovation Program for Quantum Science and Technology (Grant No. 2021ZD0300201).% put your acknowledgments here.
\end{acknowledgments}
\bibliography{qbn_with_appendix}
\onecolumngrid
\newpage
\twocolumngrid
\appendix
\section{Classical Bayesian Network}
\label{ap1}
Events with multiple values can be described as random variables. Hereafter, both events and corresponding random variables are denoted by $X$ without distinguishing them The probability distribution $P(X)$ of a random variable $X$ can be used to describe possible outcomes and probabilities of events. Multiple random variables $\{X_1, X_2, \dots, X_N\}$ have a joint probability distribution $P(X_1, X_2, \dots, X_N)$, which describes all the information of multiple events. The marginal distribution and conditional distribution can be calculated with the joint probability distribution. When there is a causal relationship between several random variables, the conditional probability distribution $P(X|Y)$ can be used to describe the relationship from the cause event $Y$ to the effect event $X$.

Bayesian network is a probabilistic graph model that describes the causal relationship and conditional probability distribution between a series of discrete type random variables. A Bayesian network is consisted of a directed acyclic graph and a series of conditional probability tables. Each node in the directed acyclic graph represents a random variable, and the directed edges denote the causal relationship between variables, where the parent nodes denote the cause events and the child nodes denote the effect events. For a general Bayesian network $\mathcal{B}$ consisting of $N$ variables, we denote the set of variables $X=\{X_i\}, i=1,2,\dots,N$, where each $X_i$ is a variable. The value of the variable $X_i$ is denoted as lowercase $x_i$. We denote the node set $\mathcal{N}$ and the edge set $\mathcal{E}=\{\mathcal{E}_{ij}\}$, where $E_{ij}$ is an edge that begins at $X_i$ and ends at $X_j$. The parent variable set and child variable set of a variable $X_i$ is defined as
\begin{align}
	&pa(X_i)=\{X_j|\mathcal{E}_{ji}\in \mathcal{E}\},\\ \nonumber
	&ch(X_i)=\{X_j|\mathcal{E}_{ij}\in \mathcal{E}\}.
\end{align}

\begin{figure}[t]
	\includegraphics[width=\linewidth, height=6cm]{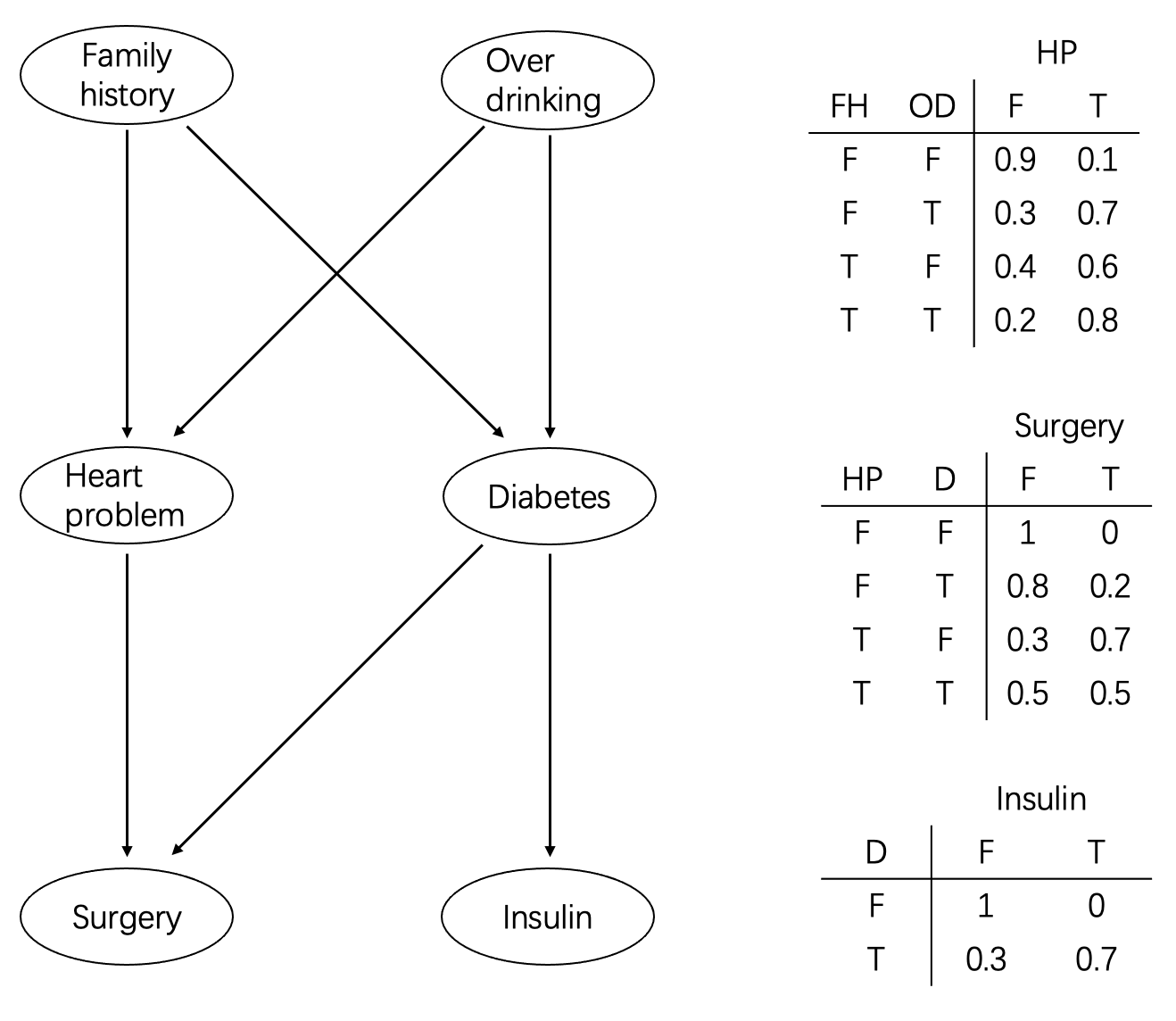}
	\caption{An example of Bayesian network. Some events in medical fields are represented as nodes, the edges represent the causal relationship between these events, and conditional probability distributions are listed as tables.}
	\label{fig4}
\end{figure}

The conditional probability tables tell us the structure of the network and the conditional probability distribution $P(X_i|pa(X_i))$ for each $X_i$. The joint probability distribution can be calculated with Eq.~(\ref{eq1}). Markovian property is an important property of Bayesian network, that is, the probability distribution of a variable in the network is only related to the value of its parent variables, but not to others. Thus,
\begin{equation}
	P(X_i|X\backslash \{X_i\})=P(X_i|pa(X_i))
\end{equation}
Based on the causal relationship between different variables, we can rewrite the joint probability distribution as a product of a series of conditional probability distributions,
\begin{equation}
	\label{eq1}
	P(X_1, X_2, \dots, X_N)=\prod_{i=1}^{N}P(X_i|pa(X_i)).
\end{equation}
Using this property, the Bayesian network only needs to store $O(N2^m)$ values rather than $2^N$ values to represent the joint probability distribution $P(X_1, X_2, \dots, X_N)$, where $m=\max_i |pa(X_i)|$ is the maximum degree of the Bayesian network. A typical Bayesian network is given in Fig~\ref{fig4}.

Inference is one of the main tasks for probabilistic models. The values of certain evidence variables $\mathcal{X}$ are observed and we want to calculate the conditional probability distribution $P(\mathcal{Y}|\mathcal{X})$ for some variables $\mathcal{Y}$ that we care about. We foucs on the inference problem on Bayesian network. In the example of Fig~\ref{fig4}, the disease a person has, i.e., the value of the evidence variables $\mathcal{X}=\{HP, D\}$, is always known. It is important for doctors to infer what is the cause of the disease. With Bayesian networks, the conditional probability distribution $P(\mathcal{Y}|\mathcal{X})$ can be calculated, where $\mathcal{Y}=\{FH, OD\}$, so that the cause of the disease can be inferred.

So, how to calculate $P(\mathcal{Y}|\mathcal{X})$ based on the Bayesian network? Bayes formula is the most commonly used tool to calculate the conditional probability distribution.
\begin{equation}
	P(\mathcal{Y}|\mathcal{X})=\frac{P(\mathcal{X}|\mathcal{Y})P(\mathcal{Y})}{P(\mathcal{X})}.
\end{equation}

\begin{figure}[h]
	\includegraphics[width=.6\linewidth]{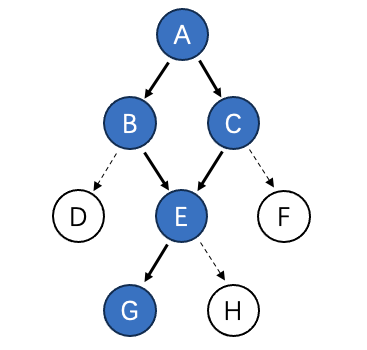}
	\caption{Suppose we want to calculate $P(G|A)$ exactly on this Bayesian network. Based on the structure of the network, $P(G|A)=\sum_{B,C,E}P(G|E)P(E|B,C)P(B|A)P(C|A)$. So exact calculation takes all path nodes(marked blue) into summation, leading to high complexity.}
	\label{fig2}
\end{figure}

Unfortunately, exact inference based on Bayes formula is proved to be $\#$P-hard~\cite{russell2016artificial}. All path nodes between $\mathcal{X}$ and $\mathcal{Y}$ need to be taken into summation in Bayes formula and the complexity is therefore exponential as shown in Fig~\ref{fig2}.

Metropolis sampling algorithm is applied instead~\cite{metropolis1953equation, chib1995understanding}. The values of the nodes are randomly sampled following the conditional probability tables. The complexity of sampling algorithm is proved to be $O(NmP(\mathcal{X}=x)^{-1})$~\cite{low2014quantum}, where $N$ is the number of nodes in the network, $m$ is the maximum indegree. As $|\mathcal{X}|$ increases, $P(\mathcal{X}=x)^{-1}$ decreases exponentially, so the sampling complexity becomes exponentially large. Other inference algorithms on Bayesian network also suffer from the problem of high computational complexity, all proved to be at least NP-hard~\cite{barkan2021representation, dagum1993approximating}.
\section{Chow-Liu Tree Structure Learning Algorithm}
\label{ap2}
The Chow-Liu algorithm is a foundational method for learning Bayesian networks from given data, which optimally approximate a joint probability distribution using pairwise dependencies~\cite{chow1968approximating}. The algorithm exploits the mutual information between the random variables and finds a Bayesian network structure for the input data, which minimizes the Kullback-Leibler (KL) divergence from the true distribution. The procedure of the algorithm is explained below.

First, a mutual information matrix containing the dependency information of the variables is calculated. The mutual information between two random variables $X$ and $Y$ describes the dependency of $X$ and $Y$, defined as
\begin{equation}
	I(X,Y)=-\frac{1}{2}\ln\left(1-Corr(X,Y)^2\right),
\end{equation}
where $Corr(X,Y)$ is the correlation function between random variables $X$ and $Y$, defined as
\begin{equation}
	Corr(X,Y)=\frac{E(XY)-E(X)E(Y)}{\sqrt{D(X)D(Y)}},
\end{equation}
where $E(X)$ is the expectation value of $X$ and $D(X)$ is the variance of $X$. Thus, $I(X,Y)$ can be calculated from the given data. For multiple random variables $\{X_1,\cdots,X_N\}$, a mutual information matrix $M$ can be defined. Each element of the mutual information matrix $M$ is defined as
\begin{equation}
	M_{ij}=I(X_i,X_j), \;\forall i,j=1,\cdots,N.
\end{equation}

Next, a maximum weight spanning tree is constructed from the mutual information matrix $M$. As the matrix $M$ is a symmetric matrix, it can be viewed as a representation of a weighted undirected graph. A maximum weight spanning tree is a connected subgraph of a weighted undirected graph that includes all vertices with no cycles and maximize the total edge weight. Kruskal's Algorithm~\cite{kruskal1956spanning} or Prim's Algorithm~\cite{prim1957shortest} can be applied. Denote a valid spanning tree $\mathcal{T}$, then the maximum weight spanning tree $\mathcal{T}^{*}$ is
\begin{equation}
	\mathcal{T}^*=\arg\max_{\mathcal{T}}\sum_{(i,j)\in \mathcal{T}} I(X_i,X_j).
\end{equation}

Finally, for each edge chosen to form the maximum weight spanning tree, the direction of the edge can be assigned arbitrarily. Thus, the spanning tree is transformed into a directed acyclic graph, which is the learned Bayesian network structure from the given data of the variables. For the convenience of the following procedure of encoding the Bayesian network into a quantum circuit, we use the depth-first search to assign the directions of the edges, such that one variable has at most one parent variable. Thus, the constructed Bayesian network has a tree structure with maximum indegree $1$. The whole Chow-Liu algorithm runs in $O(N^2)$ time where $N$ is the number of variables, making it scalable for high-dimensional data.

The reason that Chow-Liu tree structure learning algorithm works is because the KL divergence between the true joint probability distribution $P(X_1,\cdots,X_N)$ and the probability distribution of the constructed Bayesian network $Q(X_1,\cdots,X_N)$ is
\begin{equation}
	D(P\parallel Q)=-\sum_{(i,j)\in \mathcal{T}^*}I(X_i;X_j)+\sum_{i=1}^NH(X_i)-H(X_1,\ldots,X_N).
\end{equation}
Only the first term is dependent of the structure of the Bayesian network, which is minimized with the construction of the maximum weight spanning tree. Thus, the Chow-Liu algorithm learns a Bayesian network structure that minimizes the KL divergence from the true distribution.
\section{Algorithms and Theorems in the text}
\label{ap3}
We formalize the algorithms and theorems used in the QBI algorithm. The topological sorting algorithm of Bayesian network is summarized in Algorithm~\ref{alg1}. The algorithm for constructing a quantum state $|\psi\rangle$ representing a given Bayesian network $\mathcal{B}$ is summarized in Algorithm~\ref{alg2}. The theorem for performing quantum amplitude estimation with arbitrary precision and success probability is summarized in Theorem~\ref{thm1}. The theorem for performing quantum Bayesian inference with quantum amplitude estimation is summarized in Theorem~\ref{thm2}. The proof of Algorithm~\ref{alg1} is accomplished in~\cite{kahn1962topological}. The proofs of Algorithm~\ref{alg2}, Theorem~\ref{thm1}, and Theorem~\ref{thm2} are given in the subsequent sections.
\begin{algorithm}[h]
	\caption{Topological sort of graph}
	\label{alg1}
	\SetAlgoLined
	\KwIn {A directed acyclic graph $G=\langle \mathcal{N},\mathcal{E}>$, $\mathcal{N}$ is the node set and $\mathcal{E}$ is the edge set.}
	\KwOut {An array $T$, $T_i\in \mathcal{N}$. For each $i$, the indexes of the parent nodes of $T_i$ are all smaller than $i$.}
	\BlankLine
	$T\gets \varnothing$\;
	$Q\gets\{\mathcal{N}_i|\mathcal{N}_i\in \mathcal{N}, pa(\mathcal{N}_i)=\varnothing \}$\;
	\While{$Q\;!= \varnothing $}{
		remove node $N$ from the head of $Q$\;
		add node $N$ to the tail of $T$\;
		\For {$M\in ch(N)$}{
			remove the edge $(M,N)$ in $\mathcal{E}$\;
			\If {Indegree of M is $0$}{
				add node $M$ to the tail of $Q$\;
			}
		}
	}
	
	return $T$\;
\end{algorithm}

\begin{algorithm}[h]
	\caption{Construction of quantum state representing a Bayesian network}
	\label{alg2}
	\SetAlgoLined
	\KwIn {A Bayesian network $\mathcal{B}$, stored as a series of conditional probability distribution tables.}
	\KwOut {A quantum state $|\psi\rangle $. Each qubit encodes a variable, the amplitudes encode the joint probability distribution.}
	\BlankLine
	Extract graph $G=\langle \mathcal{N},\mathcal{E}\rangle$ from $\mathcal{B}$\;
	Get topologically sorted node set $T$ from algorithm \ref{alg1}\;
	Initialize a $|\mathcal{N}|$-qubit quantum state $|\psi\rangle=|00\dots0\rangle$\;
	\For{$i=1,2,\dots,|X|$}{
		\If {$pa(T_i)=\varnothing$}{
			$\theta=2\arccos\sqrt{P(T_i=0)}$\;
			apply $R_Y(\theta)$ to the $q_i$\;
		}
		\Else{
			\For{$pa(T_i)=\{0,1\}^{|pa(T_i)|}$}{
				$\theta=2\arccos\sqrt{P(T_i=0|pa(T_i))}$\;
				apply $CR_Y(\theta)$ to the $q_i$\;
			}
		}
	}
	return $|\psi\rangle$\;
\end{algorithm}

\newtheorem{theorem}{Theorem}
\begin{theorem}[Quantum amplitude estimation with arbitrary success probability]
	\label{thm1}
	The arbitrary success probability quantum amplitude estimation algorithm takes as input one copy of a quantum state $|\psi\rangle$, a unitary transformation $U=2|\psi\rangle\langle \psi|-I$, another unitary transformation $V=I-2P$ for some projector $P$, an error rate $\epsilon$, and probability of failure $\delta$. The algorithm output is $\tilde{a}$, which is an estimate of $a = \langle\psi|P|\psi\rangle$, such that
	\begin{equation}
		\frac{|\tilde{a}-a|}{|a|}\leq \epsilon
	\end{equation}
	is satisfied with probability at least $1-\delta$, using operator $UV$
	\begin{equation}
		O\left(\frac{1}{\epsilon\sqrt{a}}\ln\frac{2}{\delta}\right)
	\end{equation}
	times. If $a=0$ or $a=1$ then $\tilde{a}=a$ with certainty.
\end{theorem}
\begin{theorem}[Quantum Bayesian inference with quantum amplitude estimation]
	\label{thm2}
	The QBI algorithm takes as input a quantum state $|\psi\rangle$ representing a Bayesian network $\mathcal{B}$, a set of evidence variables $\mathcal{X}$ and target variables $\mathcal{Y}$, the value of evidence variables $\mathcal{X}=x$, an error rate $\epsilon$, and probability of failure $\delta$. The algorithm output is $\overline{P}(\mathcal{Y}|\mathcal{X}=x)$, which is an estimate of $P(\mathcal{Y}|\mathcal{X}=x)$, such that for arbitrary possible value $y$ of $\mathcal{Y}$,
	\begin{equation}
		\frac{|\overline{P}(\mathcal{Y}=y|\mathcal{X}=x)-P(\mathcal{Y}=y|\mathcal{X}=x)|}{P(\mathcal{Y}=y|\mathcal{X}=x)}\leq \epsilon
	\end{equation}
	is satisfied with probability at least $1-\delta$, using
	\begin{equation}
		O\left(N2^mP(\mathcal{X}=x)^{-\frac{1}{2}}(2\sqrt{2})^{|\mathcal{Y}|}\frac{1}{\epsilon}\ln\frac{2}{\delta}\right)
	\end{equation}
	quantum gates. $N$ is the number of variables and $m$ is the maximum indegree of $\mathcal{B}$
\end{theorem}

\section{Maximum Indegree Reduction}
\label{ap4}
It is noted that the time complexity of constructing a quantum circuit representing a Bayesian network is high. This complexity increases exponentially with the parameter $m$, the maximum indegree of Bayesian network. There is more than one kind of structure of Bayesian network that is able to represent the joint distribution $P(X_1,X_2,\dots,X_N)$ of random variables $\{X_1,X_2,\dots,X_N\}$. However, finding the globally optimized structure with minimum max-indegree is NP-hard~\cite{chickering1996learning,chickering2004large}. Instead, locally reducing indegrees of nodes is possible.

\begin{figure}[h]
	\includegraphics[width=.8\linewidth]{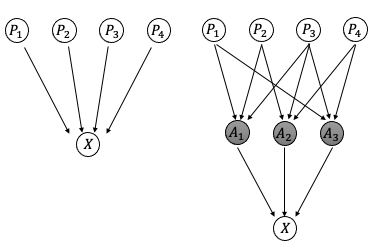}
	\caption{An example of introducing ancillary nodes.}
	\label{fig3}
\end{figure}

Each variable $X_i$ in Bayesian network contributes to the joint probability distribution with term $P(X_i|pa(X_i))$. Therefore, adding some ancillary variables between $X_i$ and $pa(X_i)$, the term $P(X_i|pa(X_i))$ can be fixed but the indegree of $X_i$ is changed. Suppose we want to add some nodes between a node $X_{j}$ and $pa(X_{j})$, where $X_{j}$ has the highest indegree in the Bayesian network. Denote the added variable set $\mathcal{A}=\{A_1,A_2,\cdots A_L\}$, each variable $A_l$ is just an ancillary node with no real-world meaning. It is required that
\begin{equation}
	P(X_j|pa(X_j))=\sum_{\mathcal{A}}P(X_j|\mathcal{A})P(\mathcal{A}|pa(X_j)),
\end{equation}
and
\begin{equation}
\label{eqa5}
	\sum_{a_l=0,1}P(A_l=a_l|pa(A_l))=1,\forall A_l \in \mathcal{A}.
\end{equation}
The size of $|\mathcal{A}|$, the structure of $\mathcal{A}$ and the conditional probability tables $P(A_l|pa(A_l))$ can be chosen manually or with machine learning methods so that the indegree of each $A_l$ and $X_j$ is smaller than the original indegree $|pa(X_j)|$. Thus, a globally NP-hard structure finding problem can be reduced to a local optimization problem.
An example of introducing ancillary nodes is shown in Fig~\ref{fig3}, requiring
\begin{align}
\label{eqa6}
	&P(X|P_1,P_2,P_3,P_4)\nonumber\\
	=&\sum_{A_1,A_2,A_3}P(X|A_1,A_2,A_3)P(A_1|P_1,P_2,P_3)\nonumber\\
	& \indent \indent \indent P(A_2|P_2,P_3,P_4)P(A_3|P_1,P_3,P_4)
\end{align}

Three hidden variables $A_1,A_2,A_3$ (marked in grey) are introduced into the original Bayesian network to suppress the maximum indegree from $4$ to $3$. The conditional probability tables $P(A_1|P_1,P_2,P_3), P(A_2|P_2,P_3,P_4), P(A_3|P_1,P_3,P_4)$ and $P(X|A_1,A_2,A_3)$ can be solved with Eq.~(\ref{eqa5}) and Eq.~(\ref{eqa6}).
\section{Previously Best Quantum Bayesian Inference Algorithm}
\label{ap5}
An algorithm in~\cite{low2014quantum} was proposed to performed Bayesian inference, which was the previously best quantum algorithm for performing Bayesian inference. First a quantum state $|\psi\rangle$ is constructed to represent a given Bayesian network. The construct procedure of $|\psi\rangle$ is same as the procedure of Algorithm~\ref{alg2}. Suppose the evidence variable set is $\mathcal{X}$ with value $\mathcal{X}=x$ and the target evidence variable set is $\mathcal{Y}$. The quantum amplitude amplification algorithm is utilized to calculate $P(\mathcal{Y}|\mathcal{X}=x)$, which is proved to achieve a quadratic speeedup over the classical sampling algorithm. We review their quantum Bayesian inference algorithm.

%When the quantum state $|\psi\rangle$ is measured in the computational basis, the value of each qubit $\{q_1,q_2,\cdots,q_N\}$ is determined as $0$ or $1$, which is a sample of the variables $\{X_1,X_2,\cdots,X_N\}$. When all the values of evidence variables in $\mathcal{X}$ satisfy $\mathcal{X}=x$, then this sample is accepted and the values of $\mathcal{Y}$ are recorded.

Suppose a quantum state $|\psi\rangle$ has been constructed, representing a given Bayesian network. $|\psi\rangle$ can be rewritten as a linear combination of the subspaces with correct evidence values and the other.
\begin{equation}
	\label{eq5}
	|\psi\rangle=\sqrt{P(\mathcal{X}=x)}|\phi\rangle|\mathcal{X}=x\rangle + \sqrt{1-P(\mathcal{X}=x)}|\overline{\phi x}\rangle
\end{equation}
Quantum amplitude amplification algorithm~\cite{grover1996fast, brassard2002quantum} is used to amplify the amplitude of state $|\phi\rangle|x\rangle$ utilizing the Grover operator, which is defined as
\begin{equation}
	\hat{G}=\hat{O}^{\dagger}S_0\hat{O}S_x,
\end{equation}
where
\begin{align}
	&\hat{O}|0\rangle=|\psi\rangle,\\\nonumber
	&S_0=I-2|0\rangle\langle 0|,\\\nonumber
	&S_x=I-2|x\rangle\langle x|.
\end{align}
Apply the Grover operator to $|\psi\rangle$ for $k$ times, where
\begin{equation}
	k\approx \lceil \frac{\pi }{4\sqrt{P(\mathcal{X}=x)^{-1/2}}}\rceil.
\end{equation}
The final state satisfies
\begin{equation}
	\hat{G}^k\hat{O}|0\rangle \approx |\phi\rangle |x\rangle
\end{equation}
with
\begin{equation}
	\parallel \langle \phi|\langle x| \;\hat{G}^k\hat{O}|0\rangle \parallel^2 =O(1).
\end{equation}
Thus, $P(\mathcal{Y}|\mathcal{X}=x)$ can be sampled from state $\hat{G}^k \hat{O}|0\rangle$ with success probabiliy $O(1)$ rather than $P(\mathcal{X}=x)$. Each use of $\hat{G}$ requires two use of $\hat{O}$, and $\hat{O}$ is implemented with $O(N2^m)$ quantum gates, where $N$ is the variable number and $m$ is the maximum indegree of the Bayesian network. So the total gate complexity for accepting one sample is $O(N2^mP(\mathcal{X}=x)^{-\frac{1}{2}})$, which has a quadratic speedup over the classical sampling complexity $O(NmP(\mathcal{X}=x)^{-1})$. Summarily, the quantum amplitude amplification based quantum Bayesian inference method increases the probability of getting a successful sample, which ultimately provides the quantum advantage.

However, to get multiple successful samples for estimating $P(\mathcal{Y}|\mathcal{X}=x)$, the algorithm in~\cite{low2014quantum} needs to be run repeatedly. The repeated run of their algorithm is a classical procedure, so there is no quantum speedup in the sampling step. To estimate $P(\mathcal{Y}=y|\mathcal{X}=x)$ for a certain $y$, according to the Hoeffding's inequality, a total number of
\begin{equation}
	\frac{1}{\epsilon^2P(\mathcal{Y}=y|\mathcal{X}=x)^2}\ln \frac{2}{\delta}
\end{equation}
samples are needed, such that
\begin{equation}
	P\left(|\frac{\overline{P}(\mathcal{Y}=y|\mathcal{X}=x)-P(\mathcal{Y}=y|\mathcal{X}=x)}{P(\mathcal{Y}=y|\mathcal{X}=x)}|<\epsilon\right)>1-\delta.
\end{equation}
Thus, to estimate all $2^{|\mathcal{Y}|}$ values of the probability distribution $P(\mathcal{Y}|\mathcal{X}=x)$, a total number of
\begin{equation}
	\frac{1}{\epsilon^2\min_yP(\mathcal{Y}=y|\mathcal{X}=x)^2}\ln \frac{2}{\delta}=O\left(\frac{4^{|\mathcal{Y}|}}{\epsilon^2}\ln \frac{2}{\delta}\right)
\end{equation}
samples are needed, such that
\begin{equation}
	P\left(|\frac{\overline{P}(\mathcal{Y}=y|\mathcal{X}=x)-P(\mathcal{Y}=y|\mathcal{X}=x)}{P(\mathcal{Y}=y|\mathcal{X}=x)}|<\epsilon\right)>1-\delta
\end{equation}
for $\forall y$.
Thus the total sampling complexity is
\begin{equation}
	O\left(N2^mP(\mathcal{X}=x)^{-\frac{1}{2}}\frac{4^{|\mathcal{Y}|}}{\epsilon^2}\ln \frac{2}{\delta}\right).
\end{equation}
Compared to the classical sampling algorithm with sampling complexity
\begin{equation}
	O\left(NmP(\mathcal{X}=x)^{-1}\frac{4^{|\mathcal{Y}|}}{\epsilon^2}\ln \frac{2}{\delta}\right).
\end{equation}
the algorithm in~\cite{low2014quantum} only achieves a quadratic speedup in the parameter $P(\mathcal{X}=x)^{-1}$, bringing an exponential term $2^m$ in the final expression.

\section{Introduction to Quantum Amplitude Estimation Algorithm}
\label{ap6}
Quantum amplitude estimation algorithm was first proposed as an application of quantum amplitude amplification algorithm~\cite{brassard2002quantum}, and later continually improved~\cite{montanaro2015quantum, giurgica2022low, grinko2021iterative}. In this section, we give a brief review to the algorithm of quantum amplitude estimation proposed in~\cite{brassard2002quantum}.

Suppose a quantum state $|\phi\rangle$ is constructed with a quantum operator $O$ as
\begin{equation}
	O|0\rangle =|\phi\rangle,
\end{equation}
and the quantum state $|\phi\rangle$ has a form of linear combination of
\begin{equation}
	|\phi\rangle=\cos\theta |x_0\rangle+\sin\theta |x_1\rangle,
\end{equation}
where $|x_0\rangle$ and $|x_1\rangle$ are two orthogonal states, and $\theta$ is an unknown angle related to the probability
\begin{equation}
	a=\sin^2\theta.
\end{equation}
The goal of the Quantum Amplitude Estimation (QAE) algorithm is to estimate $a$, i.e., the probability of measuring $|x_1\rangle$.

Defining operators
\begin{equation}
	S_0=I-2|0\rangle\langle 0|,
\end{equation}
\begin{equation}
	S_{x}=I-2|x_1\rangle\langle x_1|.
\end{equation}
Then the Grover operator is defined as
\begin{equation}
	\hat{G}=OS_0O^{\dagger}S_{x}.
\end{equation}
The operator $\hat{G}$ preserves the 2-dimensional subspace $span\{|x_0\rangle,|x_1\rangle\}$. The matrix form of $\hat{G}$ in this subspace is
\begin{equation}
	\hat{G} =
\left(
	\begin{matrix}
		\cos(2\theta) & -\sin(2\theta) \\
		\sin(2\theta) & \cos(2\theta)
	\end{matrix}
\right)
\end{equation}
with eigenvalues $e^{\pm i2\theta}$. Thus, by applying quantum phase estimation algorithm~\cite{shor1994algorithms, shor1999polynomial} to $\hat{G}$, the value of $\theta$ can be estimated. The quantum phase estimation algorithm proceeds as follows.
\begin{figure}[t]
	\includegraphics[width=1.1\linewidth]{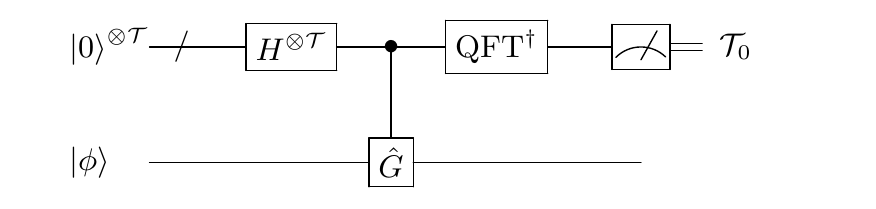}
	\caption{Quantum circuit for quantum phase estimation algorithm.}
	\label{figa4}
\end{figure}
\begin{enumerate}
	\item Two registers are used in the quantum phase estimation algorithm. An auxiliary register containing $\mathcal{T}$ qubits is initialized in state $|0\rangle^{\otimes \mathcal{T}}$, which is used for storing the estimated eigenvalues of $\hat{G}$. The auxiliary register is denoted as the first register. The second register is the data register, which is initialized in the state $|\phi\rangle$. The state of the whole system is $|0\rangle^{\otimes \mathcal{T}}|\phi\rangle$.
	\item The Hadamard gates, the controlled rotation gates and the inverse quantum Fourier transformation are applied to the state $|0\rangle^{\otimes \mathcal{T}}|\phi\rangle$ in sequence. This extracts the phase $2\theta$ and stores it in the auxiliary qubit register. The quantum circuit of quantum phase estimation is shown in Fig~\ref{figa4}.
	\item Measuring the first qubit register in the computational basis. Suppose the measure outcome is some integer $\mathcal{T}_0\in [0,2^\mathcal{T}-1]$, then the value of $\theta$ can be estimated with
	\begin{equation}
		\tilde{\theta}=\frac{\mathcal{T}_0\pi}{2^{\mathcal{T}+1}},
	\end{equation}
    Thus, the value of $a$ can be estimated with
\begin{equation}
	\tilde{a}=\sin^2\tilde{\theta}=\sin^2\left(\frac{\mathcal{T}_0\pi}{2^{\mathcal{T}+1}}\right).
\end{equation}
\end{enumerate}

Thus, the procedure gives an estimation of amplitude of $|x_1\rangle$. It is proved in~\cite{brassard2002quantum} that
\begin{equation}
	P\left(|\tilde{\theta}-\theta|<\frac{1}{2^{\mathcal{T}+1}}\right)\geq\frac{8}{\pi^2}.
\end{equation}
Thus, by enlarging the qubit number $\mathcal{T}$ in the first register, this estimation becomes exponentially accurate. We summarize the main result of~\cite{brassard2002quantum} as follows.

The quantum amplitude estimation~\cite{brassard2002quantum} takes as input one copy of a quantum state $|\psi\rangle$, a unitary transformation $U=2|\psi\rangle\langle \psi|-I$, another unitary transformation $V=I-2P$ for for some projector $P$, and an integer $m$. The algorithm output is $\tilde{a}$, which is an estimate of $a = \langle\psi|P|\psi\rangle$. If
	\begin{equation}
		\label{eq16}
		|\tilde{a}-a|\leq 2\pi \frac{\sqrt{a(1-a)}}{m}+\frac{\pi^2}{m^2}
	\end{equation}
with probability at least $8/\pi^2$, then the $m$ operator $UV$ are required. If $a=0$, then the algorithm outputs $\tilde{a}=0$ with certainty. If $a=1$ and $m$ is even, then the algorithm outputs $\tilde{a}=1$ with certainty.

However, the success probability of the quantum amplitude estimation~\cite{brassard2002quantum} is only lower bounded by a constant number of $8/\pi^2$ instead of an arbitrary value. Thus we improve the original algorithm to an arbitrary success probability and summarize our results in the main text and Theorem~\ref{thm1}.
\section{Proof of Theorem~\ref{thm1} and Theorem~\ref{thm2}}
\label{proof1}
Now we give a proof of Theorem~\ref{thm1} and Theorem~\ref{thm2}. We first prove Theorem~\ref{thm1}.
\begin{proof}
	First, we take the value of $m$ to be $m=2\lceil \pi/(\sqrt{a}\epsilon) \rceil$ as in Eq.~(\ref{eq16}). The notation $\lceil \cdot \rceil$ denotes round up. Thus
	\begin{equation}
		\begin{aligned}|\tilde{a}-a|&\leq\frac{2\pi\sqrt{a}\sqrt{a(1-a)}\epsilon}{2\pi}+\frac{\pi^{2}\epsilon^{2}a}{4\pi^{2}}\nonumber\\
&=\epsilon a\sqrt{1-a}+\frac{\epsilon^{2}a}{4}\nonumber\\
&\leq\epsilon a(1-\frac{a}{2})+\frac{\epsilon^{2}a}{4}\nonumber\\
&=\epsilon a\left(1-\frac{a}{2}+\frac{\epsilon}{4}\right)\nonumber\\
&\leq\epsilon a.\end{aligned}
	\end{equation}
Thus the quantum amplitude estimation algorithm can estimate the value of $a$ with relative error $\epsilon$ with successful probability larger than $8/\pi^2$. Suppose quantum amplitude estimation algorithm are used for multiple times to get multiple estimations $a_1,a_2,\cdots,a_J$ of $a$. Then $a_1,a_2,\cdots,a_J$ is independent with each other, satisfying the same probability distribution. Each of the $a_j$ satisfies
\begin{equation}
	P\left(\frac{|a_j-a|}{|a|}\leq\epsilon\right)\geq \frac{8}{\pi^2},
\end{equation}
and
\begin{equation}
	E(a_j)=a,\forall j=1,2,\cdots, J.
\end{equation}
According to the Chebyshev's inequality
\begin{equation}
	P(|a_j-E(a_j)|\leq\epsilon)\geq 1-\frac{\sigma^2}{\epsilon^2},\forall j=1,2,\cdots, J,
\end{equation}
where $\sigma^2$ is the variance of each $a_j$.
Thus,
\begin{equation}
	1-\frac{\sigma^2}{a^2\epsilon^2}\geq\frac{8}{\pi^2},
\end{equation}
giving us an upper bound of the variance of each $a_j$ as
\begin{equation}
	\label{eq8}
	\sigma^2\leq a^2\epsilon^2\left(1-\frac{8}{\pi^2}\right).
\end{equation}
Denote $\tilde{a}$ to be an estimation of $a$ with
\begin{equation}
	\tilde{a}=\frac{1}{J}\sum_{j=1}^{J}a_j.
\end{equation}
Then according to the law of large numbers, $\tilde{a}$ obeys normal distribution $f(\tilde{a})$
\begin{equation}
	\tilde{a}\sim f(\tilde{a})=N\left(a,\frac{\sigma^2}{J}\right).
\end{equation}
Letting
\begin{equation}
	\label{eq9}
	\int_{a-a\epsilon}^{a+a\epsilon} f(\tilde{a})d\tilde{a}\geq 1-\delta.
\end{equation}
Combining Eq.~(\ref{eq8}) and Eq.~(\ref{eq9}), an lower bound of $J$ is given as
\begin{equation}
	\label{eq10}
	J\geq \frac{2(\pi^2-8 )(Erf^{-1}(1-\delta))^2}{\pi^2},
\end{equation}
where $Erf^{-1}(x)$ is the inverse function of $Erf(x)$ which is defined as
\begin{equation}
	Erf(x)=\frac{2}{\sqrt{\pi}}\int_{t=0}^{x}e^{-t^2}dt.
\end{equation}
Expanding the $Erf^{-1}(1-\delta)$ at $\delta=0$:
\begin{equation}
	2(Erf^{-1}(1-\delta))^2\approx \ln\frac{2}{\pi \delta^2}
\end{equation}
Thus, taking
\begin{equation}
	J=O\left(\ln\frac{2}{\delta}\right),
\end{equation}
Eq.~(\ref{eq10}) can be satisfied. Summarily, if we perform $J$ times of quantum amplitude estimation of $a$ to get $a_1,a_2,\cdots,a_J$ with $J=O(\ln 2/\delta)$ and each estimation approximates $a$ to accuracy $\epsilon$ with probability larger than $8/\pi^2$, then $\tilde{a}=\sum_{j=1}^{J}a_j/J$ is an estimation of $a$, and $|\tilde{a}-a|/|a|<\epsilon$ is satisfied with probability larger than $1-\delta$. The total uses of operator $UV$ as defined in the quantum amplitude algorithm is then
\begin{equation}
	O\left(\frac{1}{\epsilon\sqrt{a}}\ln\frac{2}{\delta}\right).
\end{equation}
As we take the number $m$ in the quantum amplitude estimation to be $2\lceil \pi/(\sqrt{a}\epsilon) \rceil$, which is an even number. Then following the results of~\cite{brassard2002quantum}, if $a=0$ or $a=1$, then the algorithm outputs $\tilde{a}=a$ with certainty.
Thus, Theorem~\ref{thm1} is proved.
\end{proof}

Next, Theorem~\ref{thm2} is proved.
\begin{proof}
	To estimate $P(\mathcal{Y}=y|\mathcal{X}=x)$ for a certain $y$, following the Bayesian formula
	\begin{equation}
		P(\mathcal{Y}=y|\mathcal{X}=x)=\frac{P(\mathcal{Y}=y,\mathcal{X}=x)}{P(\mathcal{X}=x)},
	\end{equation}
we need to estimate the value of $P(\mathcal{Y}=y,\mathcal{X}=x)$ and $P(\mathcal{X}=x)$. Following Theorem~\ref{thm1}, when the projector is taken as
\begin{equation}
	P=I\otimes |\mathcal{X}=x\rangle \langle \mathcal{X}=x|,
\end{equation}
$P(\mathcal{X}=x)$ is estimated. When the projector is taken as
\begin{equation}
	P=I\otimes |\mathcal{Y}=y,\mathcal{X}=x\rangle \langle \mathcal{Y}=y,\mathcal{X}=x|,
\end{equation}
$P(\mathcal{Y}=y,\mathcal{X}=x)$ is estimated.
Suppose for a certain $y$, we use Theorem~\ref{thm1} to estimate $P(\mathcal{Y}=y,\mathcal{X}=x)$ to relative error $\epsilon/3$ with probability larger than $1-\delta/2$. And $P(\mathcal{X}=x)$ is estimated to relative error $\epsilon/3$ with probability larger than $1-\delta/2$. It is required that $\epsilon<1/3$. Then a total number of
\begin{equation}
	O\left(\frac{1}{\epsilon\sqrt{P(\mathcal{Y}=y,\mathcal{X}=x)}}\ln\frac{2}{\delta}\right)
\end{equation}
operators $UV$ are used. Denote $P(\mathcal{Y}=y,\mathcal{X}=x)=P_{xy}$ and the estimation of $P_{xy}$ is $\overline{P_{xy}}$. Denote $P(\mathcal{X}=x)=P_{x}$ and the estimation of $P_{x}$ is $\overline{P_{x}}$, then
\begin{equation}
	P\left(|\overline{P_{xy}}-P_{xy}|\leq\frac{\epsilon P_{xy}}{3}\right)\geq1-\frac{\delta}{2},
\end{equation}
\begin{equation}
	P\left(|\overline{P_{x}}-P_{x}|\leq \frac{\epsilon P_{x}}{3}\right)\geq 1-\frac{\delta}{2}.
\end{equation}
When both $|\overline{P_{xy}}-P_{xy}|\leq \epsilon P_{xy}/3$ and $|\overline{P_{x}}-P_{x}|\leq\epsilon P_{x}/3$ hold true, then
\begin{equation}
	\frac{|\overline{P_{xy}}/\overline{P_{x}}-P_{xy}/P_x|}{|P_{xy}/P_x|} \leq\frac{-2\epsilon}{3-\epsilon}<\epsilon.
\end{equation}
Thus,
\begin{equation}
	P\left(\frac{|\overline{P_{xy}}/\overline{P_{x}}-P_{xy}/P_x|}{|P_{xy}/P_x|} <\epsilon\right)\geq (1-\frac{\delta}{2})^2>1-\delta
\end{equation}
is proved, which is
	\begin{equation}
		P\left(|\frac{\overline{P}(\mathcal{Y}=y|\mathcal{X}=x)-P(\mathcal{Y}=y|\mathcal{X}=x)}{P(\mathcal{Y}=y|\mathcal{X}=x)}|<\epsilon\right)>1-\delta.
	\end{equation}
Therefore, to estimate $P(\mathcal{Y}=y|\mathcal{X}=x)$ to relative error $\epsilon$ with probability larger than $1-\delta$, the total number of uses of operators $UV$ as defined in Theorem~\ref{thm1} is
\begin{equation}
	O\left(\frac{1}{\epsilon\sqrt{P(\mathcal{Y}=y,\mathcal{X}=x)}}\ln\frac{2}{\delta}\right).
\end{equation}
	Thus, to estimate all $2^{|\mathcal{Y}|}$ values of the probability distribution $P(\mathcal{Y}|\mathcal{X}=x)$, a total number of
	\begin{equation}
			O\left(\frac{2^{|\mathcal{Y}|}}{\epsilon\min_y\sqrt{P(\mathcal{Y}=y,\mathcal{X}=x)}}\ln\frac{2}{\delta}\right).
	\end{equation}
	uses of operators $UV$ are needed, such that
	\begin{equation}
		P\left(|\frac{\overline{P}(\mathcal{Y}=y|\mathcal{X}=x)-P(\mathcal{Y}=y|\mathcal{X}=x)}{P(\mathcal{Y}=y|\mathcal{X}=x)}|<\epsilon\right)>1-\delta
	\end{equation}
	for $\forall y$. And
	\begin{equation}
		\min_y\sqrt{P(\mathcal{Y}=y,\mathcal{X}=x)}\geq P(\mathcal{X}=x)^{-1/2} \sqrt{2}^{-|\mathcal{Y}|}.
	\end{equation}
Thus, the total number of uses of operators $UV$ is
\begin{equation}
	O\left(P(\mathcal{X}=x)^{-1/2} \frac{(2\sqrt{2})^{|\mathcal{Y}|}}{\epsilon}\ln\frac{2}{\delta}\right).
\end{equation}

$U$ and $V$ are defined as
	\begin{equation}
		U=2|\psi\rangle\langle \psi|-I
	\end{equation}
and
\begin{equation}
	V=I-2P.
\end{equation}
The gate complexity of one use of $U$ is $O(N2^m)$ while the gate complexity of one use of $V$ is $O(N)$, where $N$ is the variable number and $m$ is the maximum indegree of the original Bayesian network. Thus, the total gate complexity for estimating $P(\mathcal{Y}|\mathcal{X}=x)$ is
	\begin{equation}
		O\left(N2^mP(\mathcal{X}=x)^{-1/2} \frac{(2\sqrt{2})^{|\mathcal{Y}|}}{\epsilon}\ln\frac{2}{\delta}\right).
	\end{equation}
Consequently, Theorem~\ref{thm2} is proved.
\end{proof}
\onecolumngrid
\section{Proof of Algorithm~\ref{alg2}}
\label{proof2}
\begin{proof}
	We use mathematical induction method to prove the correctness of Algorithm~\ref{alg2}.
    First, for $k=1$, following Algorithm~\ref{alg2}, a quantum state $|\psi_1\rangle$ is constructed with a rotation-y quantum gate $\rm{R_Y}(\theta)$ where $\theta=2\arccos P(X_1=0)$.
    \begin{equation}
    |\psi_1\rangle=\sqrt{P(X_1=0)}|0\rangle+\sqrt{1-P(X_1=0)}|1\rangle=\sqrt{P(X_1=0)}|0\rangle+\sqrt{P(X_1=1)}|1\rangle
    \end{equation}
    Thus, Algorithm~\ref{alg2} is correct for the case $k=1$.

	Suppose for the first $k$ variables, we have successfully construct a quantum state $|\psi_{k}\rangle$ to represent the joint probability distribution
	\begin{equation}
		\label{eqa1}
		|\psi_k\rangle =\sum_{q_1,\cdots,q_{k}}\sqrt{P(X_1=q_1,\cdots,X_k=q_k)}|q_1q_2\cdots q_k\rangle.
	\end{equation}
	Denote
	\begin{equation}
		K=\{1,2,\cdots,k\}
	\end{equation}
	$S$ is the index of $pa(X_{k+1})=\{X_{s_1},,X_{s_2},\cdots,X_{s_m}\}$
	\begin{equation}
		S=\{s_1,s_2,\cdots,s_m\}.
	\end{equation}
    Algorithm~\ref{alg2} utilizes $2^m$ quantum gates $U_1,U_2,\cdots,U_{2^m}$ to build state $|\psi_{k+1}\rangle$ based on state $|\psi_k\rangle$ as
	\begin{equation}
		|\psi_{k+1}\rangle=\prod_{j=1}^{2^m}U_j|\psi_{k}\rangle|0\rangle.
	\end{equation}
Each of the $U_j$ is a controlled rotation-y operator, where the control qubits are the qubits representing $pa(X_{k+1})$ and the target qubit is the $(k+1)$th qubit which represents variable $X_{k+1}$.
	We now prove the constructed $|\psi_{k+1}\rangle$ satisfies
	\begin{equation}
		\label{eqa2}
		|\psi_{k+1}\rangle=\sum_{q_1,\cdots,q_{k},q_{k+1}}\sqrt{P(X_1=q_1,\cdots,X_k=q_k,X_{k+1}=q_{k+1})}|q_1q_2\cdots q_kq_{k+1}\rangle.
	\end{equation}

    Combining Eq.~(\ref{eqa1}) and Eq.~(\ref{eqa2}) we have
	\begin{equation}
		|\psi_{k+1}\rangle=\prod_{j=1}^{2^m}U_j\sum_{q_1,\cdots,q_{k}}\sqrt{P(X_1=q_1,\cdots,X_k=q_k)}|q_1q_2\cdots q_k\rangle|0\rangle.
	\end{equation}
    Change the order of production and summation, and divide the qubits that are taken into summation into two sets: the qubits represent $pa(X_{k+1})$ and others,
	\begin{equation}
		|\psi_{k+1}\rangle=\sum_{q_i,i\in K\backslash S}\sum_{q_i,i\in S}\prod_{j=1}^{2^m}U_j \sqrt{P(X_1=q_1,\cdots,X_k=q_k)}|q_1q_2\cdots q_k\rangle|0\rangle.
	\end{equation}
Applying $\prod_{j=1}^{2^m}U_j$ to state $|q_1q_2\cdots q_k\rangle|0\rangle$ gives us
	\begin{equation}
		\label{eqa3}
		|\psi_{k+1}\rangle=\sum_{q_i,i\in K\backslash S}\sum_{q_{s_1},\cdots,q_{s_m}} \sqrt{P(X_1=q_1,\cdots,X_k=q_k)}|q_1q_2\cdots q_k\rangle\left(\alpha_{q_{s_1}\cdots q_{s_m}}|0\rangle +\beta_{q_{s_1}\cdots q_{s_m}}|1\rangle\right),
	\end{equation}
where according to Algorithm~\ref{alg2},
	\begin{equation}
		\alpha_{q_{s_1}\cdots q_{s_m}}=\sqrt{P\left(X_{k+1}=0|X_{s_1}=q_{s_1},\cdots,X_{s_m}=q_{s_m}\right)},
	\end{equation}
	\begin{equation}
		\beta_{q_{s_1}\cdots q_{s_m}}=\sqrt{P\left(X_{k+1}=1|X_{s_1}=q_{s_1},\cdots,X_{s_m}=q_{s_m}\right)}.
	\end{equation}
Thus,
	\begin{align}
		P(X_1=q_1,\cdots,X_k=q_k,X_{k+1}=0)&=P(X_{k+1}=0|X_1=q_1,\cdots,X_k=q_k)P(X_1=q_1,\cdots,X_k=q_k)\nonumber\\&=P\left(X_{k+1}=0|X_{s_1}=q_{s_1},\cdots,X_{s_m}=q_{s_m}\right)P(X_1=q_1,\cdots,X_k=q_k)\nonumber\\&= \alpha^2_{q_{s_1}\cdots q_{s_m}}P(X_1=q_1,\cdots,X_k=q_k),
	\end{align}
	\begin{align}
		P(X_1=q_1,\cdots,X_k=q_k,X_{k+1}=1)&=P(X_{k+1}=1|X_1=q_1,\cdots,X_k=q_k)P(X_1=q_1,\cdots,X_k=q_k)\nonumber\\&=P\left(X_{k+1}=1|X_{s_1}=q_{s_1},\cdots,X_{s_m}=q_{s_m}\right)P(X_1=q_1,\cdots,X_k=q_k)\nonumber\\&= \beta^2_{q_{s_1}\cdots q_{s_m}}P(X_1=q_1,\cdots,X_k=q_k).
	\end{align}
Combining the two equations above and Eq.~(\ref{eqa3}),
	\begin{align}
		|\psi_{k+1}\rangle=&\sum_{q_i,i\in K\backslash S}\sum_{q_{s_1},\cdots,q_{s_m}} \sqrt{P(X_1=q_1,\cdots,X_k=q_k,X_{k+1}=0)}|q_1q_2\cdots q_k0\rangle\nonumber\\
		&+\sqrt{P(X_1=q_1,\cdots,X_k=q_k,X_{k+1}=1)}|q_1q_2\cdots q_k1\rangle,
	\end{align}
which is
	\begin{equation}
		|\psi_{k+1}\rangle=\sum_{q_1,\cdots,q_{k},q_{k+1}}\sqrt{P(X_1=q_1,\cdots,X_k=q_k,X_{k+1}=q_{k+1})}|q_1q_2\cdots q_kq_{k+1}\rangle.
	\end{equation}
It is shown that $|\psi_{k+1}\rangle$ represents the joint probability distribution of the first $k+1$ variables. Thus, we prove the correctness of Algorithm~\ref{alg2} with the mathematical induction method.
\end{proof}

\section{Examples of Wafer Bin Maps}
\label{ap9}
Some examples of different defect types on wafer bin maps. There are total nine kinds of defect types on the wafer bin maps: Normal, Center, Donut, Edge-Loc, Edge-Ring, Loc, Near-full, Scratch, Random. Some examples are shown.

The first column has the types of the defects. The second column is the original data from the data set we use. The third column is the bivalued original data. The fourth column shows the compressed data. The original data as in the second column has three different values: 0, 1, and 2. 1 represents that there is no error on the chip, marked in green. 2 represents that there is an error on this chip, marked in yellow. 0 represents that there is no chip at this pixel, marked in purple. For convenience of Bayesian network construction, we first do data preprocess to turn the original data into bivalued data with only 0 and 1. The bivalued data is shown in the third column, where the purple pixels represent data 0 and the yellow pixels represent data 1. Finally, major voting compression methods are used to compress the data from 52-by-52 to 8-by-8. The compressed data is shown in the last column where the purple pixels represent data 0 and the yellow pixels represent data 1.
\newpage
\begin{figure}[H]
	\centering
	\includegraphics[width=.8\linewidth]{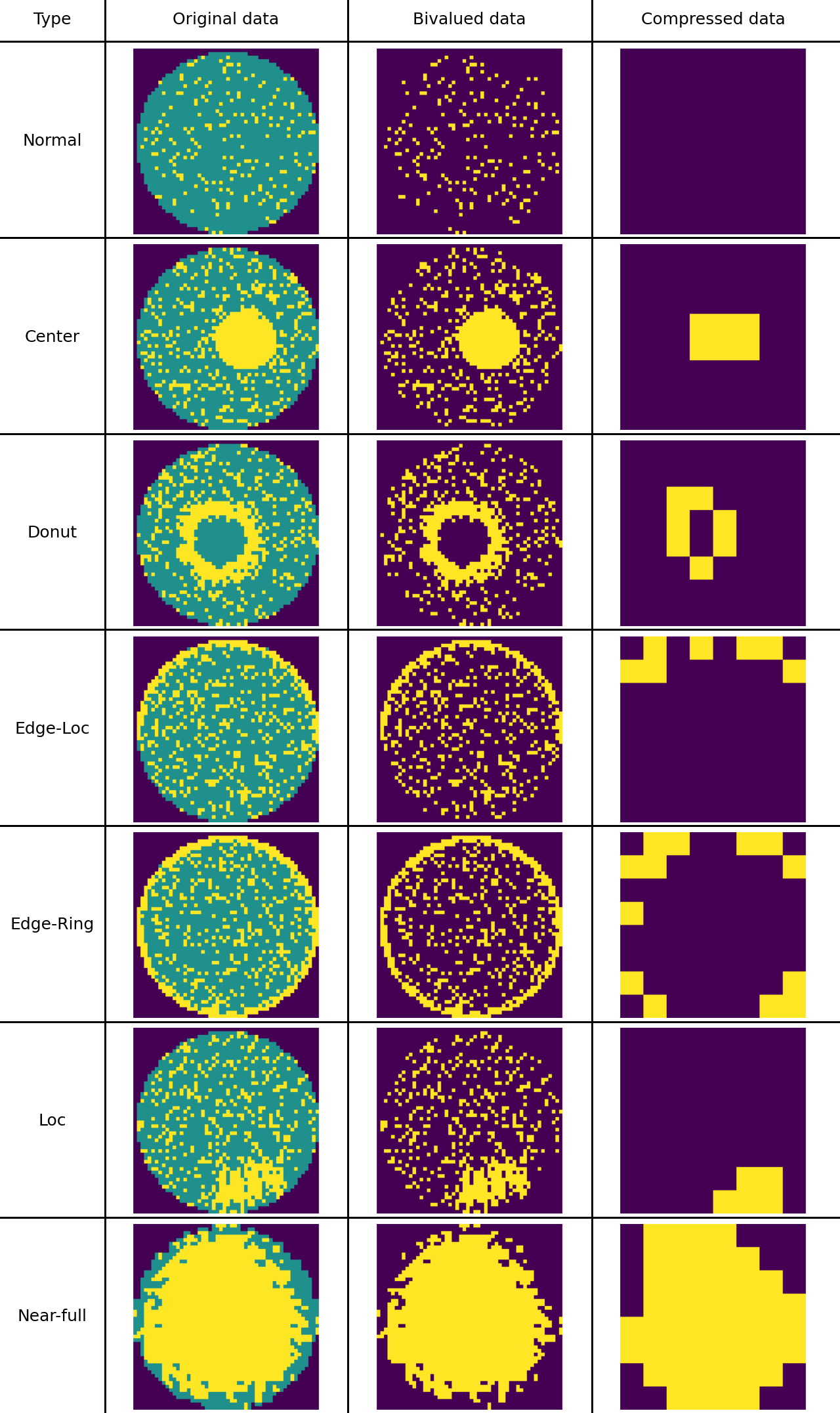}\\
 \label{fig51}
\end{figure}
\newpage
\begin{figure}[H]
	\centering
	\includegraphics[width=.8\linewidth]{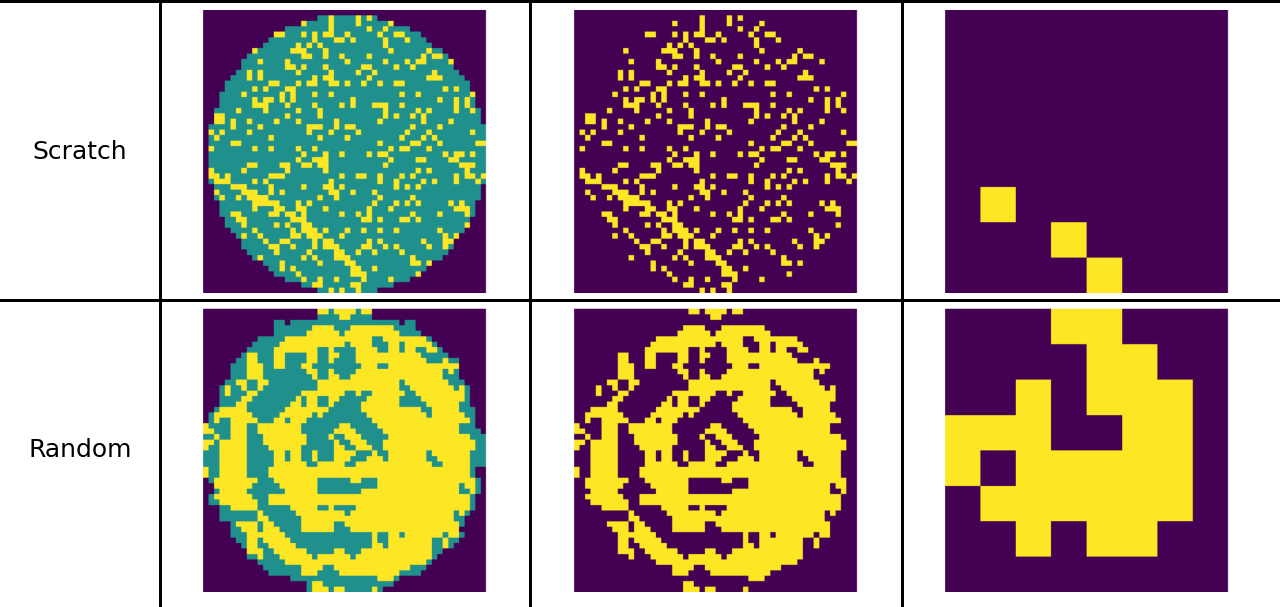}\\
\label{fig52}
\end{figure}

\section{Confusion Matrices of Training and Testing}
\label{ap10}
\begin{figure}[tbph]
 	\centering
 	\subfigure {\
 		\begin{minipage}[b]{.45\linewidth}
 			\centering
 			\begin{overpic}[scale=0.45]{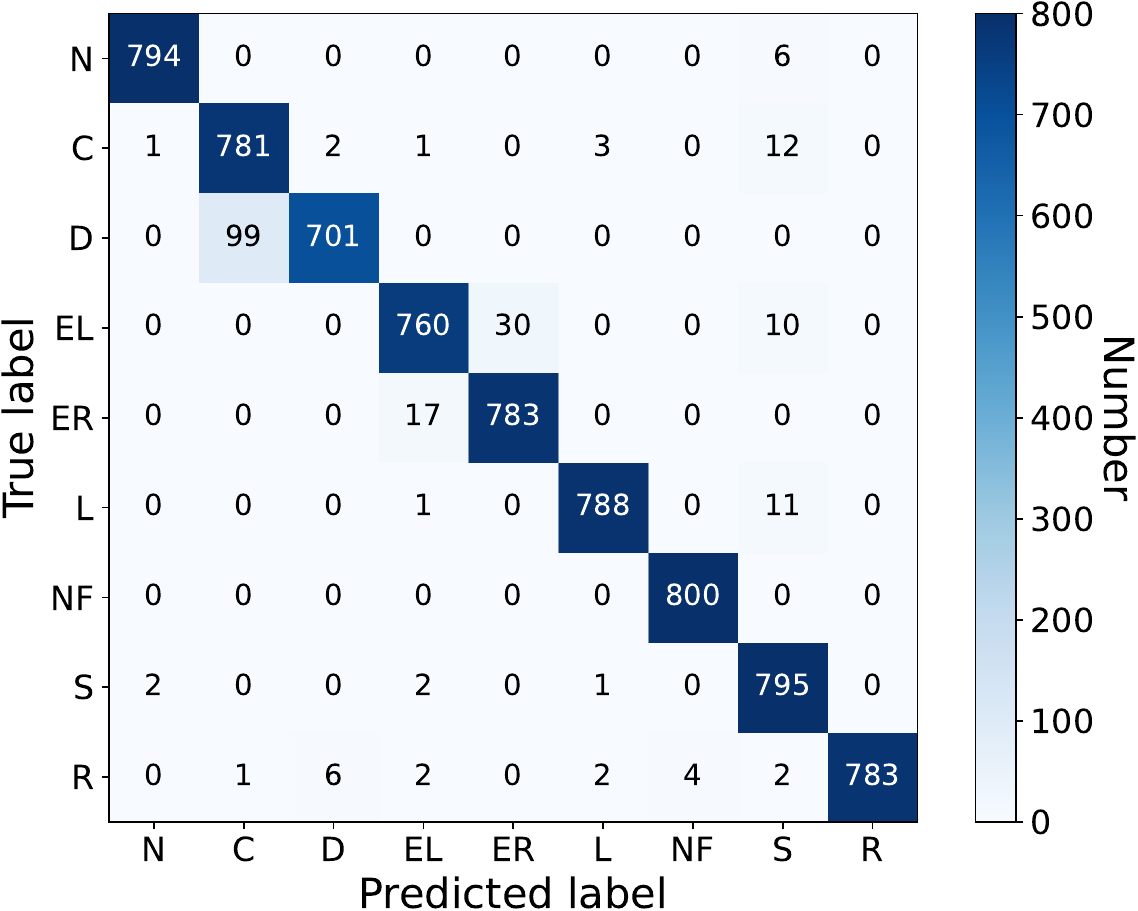}
 				\put(0,80){\large\textbf{(a)}}
 			\end{overpic}
 		\end{minipage}
 	}\;\;\;\;\;\;
 	\subfigure {\
 		\begin{minipage}[b]{.45\linewidth}
 			\centering
 			\begin{overpic}[scale=0.45]{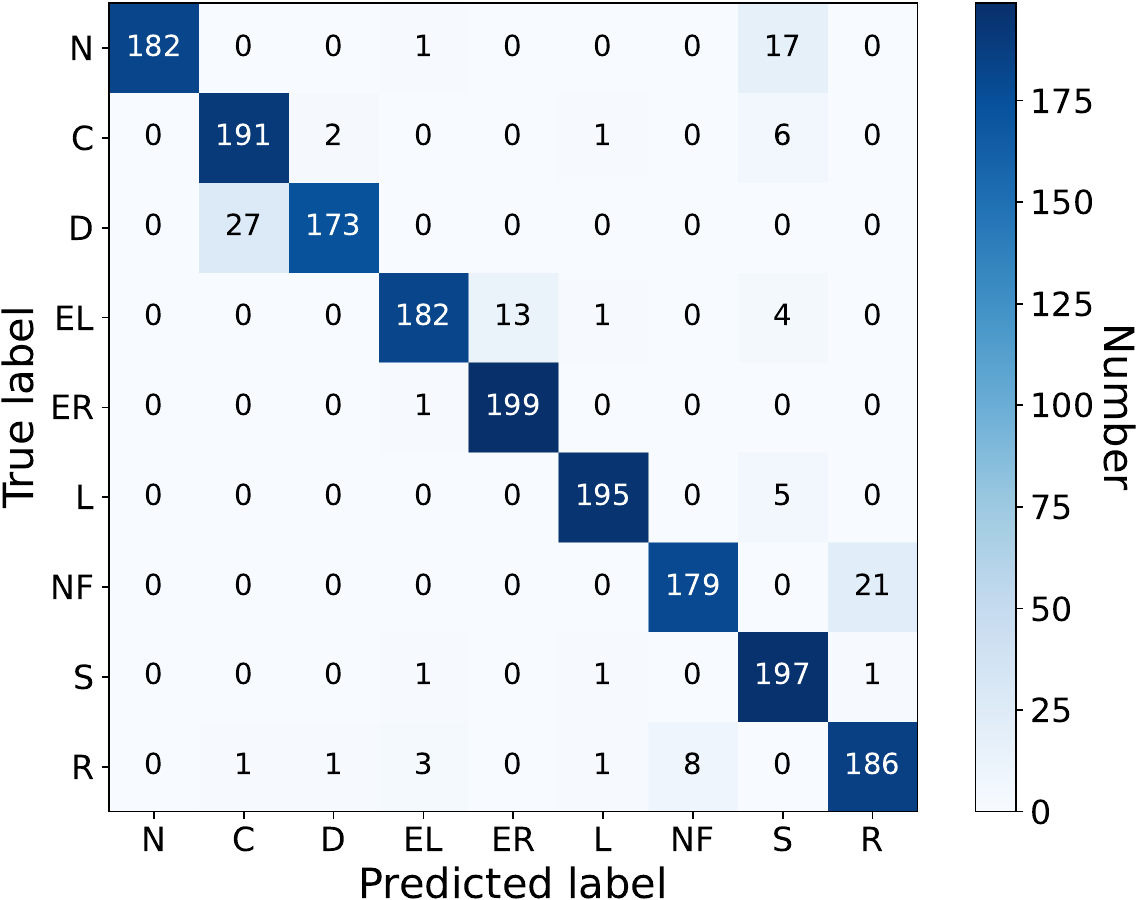}
 				\put(0,80){\large\textbf{(b)}}
 			\end{overpic}
 		\end{minipage}
 	}
 \caption{(a) The confusion matrix of quantum Bayesian inference on the training set. (b) The confusion matrix of quantum Bayesian inference on the test set.}
 \label{figs6}
 \end{figure}

 The confusion matrix of quantum Bayesian inference on the training set is shown in Fig.~\ref{figs6}(a). The confusion matrix of quantum Bayesian inference on the test set is shown in Fig.~\ref{figs6}(b). Each element in the confusion matrix represents the number of data with the correct category as the horizontal index and the classification result as the vertical index.  The characters on the axis represent the type of defect. N stands for Normal. C stands for Center. D stands for Donut. EL stands for Edge-Loc. ER stands for Edge-Ring. L stands for Loc. NF stands for Near-Full. S stands for Scratch. R stands for Random. The data on the diagonal is classified correctly, and the data on the non-diagonal is classified incorrectly. The value of diagonal elements is much larger than that of non-diagonal elements as shown in Fig.~\ref{figs6}(a) and Fig.~\ref{figs6}(b), and only similar categories may cause misclassification. This proves the reliability and high accuracy of the quantum Bayesian inference method.

\end{document}